\documentclass{emulateapj}

\accepted{2006 July 10}
\received{2006 April 10}
\shorttitle{Galaxy Luminosity Function at $0.03<z<0.5$}

\shortauthors{Xia et al.}

\begin{document}

\title{Evolution of Galaxy Luminosity Function and Luminosity Function by Density Environment
at $0.03<z<0.5$}

\author{
Lifang Xia\altaffilmark{1,2}, Xu Zhou\altaffilmark{1},
Yanbin Yang\altaffilmark{1}, Jun Ma\altaffilmark{1},
Zhaoji Jiang\altaffilmark{1}}

\email{xlf@vega.bac.pku.edu.cn}
\email{zhouxu@vega.bac.pku.edu.cn}

\altaffiltext{1}{National Astronomical Observatories,
Chinese Academy of Sciences, Beijing, 100012, P. R. China}
\altaffiltext{2}{Department of Astronomy, Peking University,
Beijing, 100871, P. R. China}

\begin{abstract}

Using galaxy sample observed by the BATC large-field multi-color sky survey and galaxy data of SDSS
in the overlapped fields, we study the dependence of the restframe $r$-band galaxy luminosity function
on redshift and on large-scale environment. The large-scale environment is defined by isodensity
contour with density contrast $\delta\rho / \rho$. The data set is a composite sample of
69,671 galaxies with redshifts
$0.03<z<0.5$ and $r<$ 21.5 mag. The redshifts are composed by three parts: 1) spectroscopic
redshifts in SDSS for local and most luminous galaxies; 2) 20-color photometric redshifts derived
from BATC and SDSS; 3) 5-color photometric redshifts in SDSS. We find that the faint-end slope
$\alpha$ steepens slightly from $-1.21$ at $z \sim 0.06$
to $-1.35$ at $z \sim 0.4$, which is the natural consequence of the hierarchical formation of galaxies.
The luminosity function also differs with different environments.
The value of $\alpha$ changes from $-1.21$
at underdense regions to $-1.37$ at overdense regions and the corresponding $M_{\ast}$
brightens from $-22.26$ to $-22.64$.
This suggests that the fraction of faint galaxies is larger in high density regions than in low density
regions.

\end{abstract}


\section{INTRODUCTION}

Galaxy luminosity function (LF) is a powerful tool in the study of galaxy formation and evolution. Galaxy LF
is directly related to galaxy mass function. Press and Schechter (1974) present a simple analytical formula
for the mass distribution based on the hierarchical assembly of galaxies. \citet{schechter76} gives an emperical
functional form for galaxy LF. It has been proposed that it should be universal \citep{lugger86,colless89,trentham98}.
However, mass function, star-formation process as well as morphological characteristics of galaxies are affected
by their environments and evolve with time, galaxy LF is expected to change with time
and to vary with galaxy characteristics and the density environments.
Deep wide-area sky surveys, such as the Two-Degree Field Galaxy Redshift Survey
\citep[2dFGRS;][]{colless01,norberg02} and Sloan Digital Sky Survey \citep[SDSS;][]{york00},
can generate large samples with a range of redshifts, which are best suited for the mesurment of LF of galaxies.
Many studies have been done in galaxy LF by morphological types \citep{wolf03,cross04,croton04},
by redshifts \citep{lilly95,ellis96,wolf03,loveday04}
and by large-scale environments \citep{mercurio03,haines04,croton04,hoyle03}.
These enable us to test the theories of formation and evolvement of galaxies
in different cosmological models.

\citet{ellis96} indicate that there is indeed a steepening in the faint-end slope with redshift. Limited by the
size of data sets, however, the evolution of LF to high redshift was not
constrained well. For galaxy LF by large-scale environment, analyses reveal that,
at low-density environment \citep{efstathiou88,loveday92,hoyle03},
the faint-end slope turns out to be $\alpha\sim -1$, and at high-density regions the
slope seems to be steeper with $-1.8<\alpha<-1.3 $ \citep{de95,lumsden97,valotto97}.

In this paper, we use our BATC (Beijing-Arizona-Taiwan-Connecticut) 15 intermediate-band
color sky survey data and SDSS 5-color sky survey data to study the evolution of LF and the environmental
effects on the LF. BATC photometric system has an average depth of 20.5 mag and the corresponding
redshifts of most galaxies are less than 0.3.  The
SDSS broadband photometric system has an average depth of 23.0 mag with most redshifts less
than 0.5. From the investigation in \citet{xia02},
the accuracy of redshift determination by 15 intermediate-bands, $\sigma_{z}\sim0.02$,
is much better than that of broadband, $\sigma_{z}\sim0.05$, in the same photometric magnitude errors.
Therefore, with the accurate photometric redshifts at $z<0.3$ and with small ${\Delta z}\over {z}$
at $0.3<z<0.5$, it is possible to combine data of these two systems for
accurate measurements of galaxy LF to high redshifts.

The content of this paper is as follows. In $\S$ 2 we briefly describe the data sample, the
application of photometric redshift code $hyperz$ and $k$-correction.
The fitting of galaxy LF and results of LF evolution are given in $\S$ 3.
Method of environmental classification and results of environmental effects on LF are presented
in $\S$ 4. $\S$ 5 discusses the effects of photometric redshift uncertainty on galaxy LF shape
by simulation and summarizes our conclusions. Through out this paper, we assume a
$\Lambda$CDM cosmological model with matter density $\Omega_{m}= 0.3$, vacuum density $\Omega_{\Lambda}= 0.7$, and
Hubble constant $H_{0}= 100h\rm{kms^{-1} Mpc^{-1}}$ with $h= 0.75$ for the calculation of distances and
volumes \citep[see][]{hogg99}.

\section{DATA}

\subsection {Sample}

We use galaxy data including BATC 15-color photometries, SDSS spectroscopies and 5-colors photometries.
The BATC Sky Survey performs photometric observations with a large field multi-color system. The observation is carried
out with the 60/90 cm f/3 Schmidt Telescope of National Astronomical Observatories, Chinese Academy of Sciences, (NAOC)
located at the Xinglong station. For detailed description of survey and performance, see Zhou
et al. 2002. The SDSS performs imaging and spectroscopic surveys over $\pi$
steradians in the northern Galactic cap with a 2.5 m telescope at Apache Point Observatory, Sunspot, New Mexico
\citep{york00}. The detailed description of the photometric and spectroscopic parameters can be found
in \citet{stoughton02}.

We select 17 fields, totally $\sim$ 17 deg$^2$, observed by BATC and overlapped
with SDSS sky survey. Galaxies with the spectroscopic and photometric information in SDSS are obtained from
SDSS Data Release 2 (http://www.sdss.org/dr2/). The data of 69,671 galaxies are achieved in SDSS
with $r<$ 21.5 and $0.03<z<0.5$.
Galaxies in BATC are selected by coordinates given by BATC and SDSS.
10,681 galaxies in BATC are obtained with the distance deviations in BATC and in SDSS less than
$2^{\prime\prime}{\mbox{}\hspace{-0.1cm}.0}$. To combine photometries in BATC
and SDSS, we need to apply aperture correction to SDSS model magnitudes since that an aperture of 4 pixels
(i.e., $r_{\rm ap}$= $6^{\prime\prime}{\mbox{}\hspace{-0.1cm}.8}$) is applied in BATC photometries
\citep[see details from][]{yuan03}. The formula of aperture correction is as below:
\begin{equation}
\Delta m=m_{\rm ap}-m_{\rm model}=-2.5\log\frac {\int_{0}^{r_{\rm ap}}2\pi rI(r)dr} {\int_{0}^{\infty}2\pi rI(r)dr}
\end{equation}
where $m_{\rm ap}$ is the aperture magnitude, $I(r)$ is the profile function of surface intensity for
the best fit of de Vaucouleurs or exponential model. For these common galaxies,
we estimate photometric redshifts by the total 20 colors.

\subsection{Redshift}

The redshifts in our catalog are measured by three methods. 1,362
galaxies have spectroscopic redshifts observed by SDSS;
10,681 galaxies have photometric redshifts estimated by 20 color photometries consisting of BATC and SDSS;
and the rest 57,628 galaxies have photometric redshifts estimated
by 5 color photometries of SDSS. The photometric redshift technique is based on SED fitting to estimate
redshifts by comparing the spectrum of an object, which should include several strong spectral features
such as 4000{\AA} break, Lyman-forest decrement, etc, with the template spectra.
We use the $hyperz$ program developed by \citet{bolzonella00} to estimate redshifts.
The accuracy of redshift determination in BATC has been achieved to be 0.02 and the estimated accuracy by
broadband filters is about 0.05 \citep{xia02}.

The accuracy of photometric redshifts is assessed by galaxies with
spectroscopic redshifts in the fields for both systems. Fig. 1
shows the comparision between photometric redshifts and
spectroscopic redshifts. We can distinctly see that redshifts of
certain amount galaxies are overestimated by the 5-color SDSS
photometries. The combined 20 colors can take them back to
reasonable estimation. The redshifts in the range from 0.3 to 0.4
seem totally to be overestimated in SDSS, and this may effect the
measure of luminosity function dramatically because of the
dominant amount of galaxies of SDSS in high redshift layer. The
uncertainties are $\sigma$= 0.017 for 20-color galaxies and
$\sigma$= 0.022 for 5-color galaxies with excluding those $\Delta
z>$ 0.05. Considered that galaxies used to estimate accuracy are
bright in magnitudes and accurate in photometries, for faint
galaxies, the errors of estimated photometric redshifts should be
larger than these estimates. To assess how the uncertainty of
photometric redshift affect the predicted luminosity function, we
will investigate by simulation in $\S$ 5s. Galaxies with
$z<0.03$ are excluded in the construction of galaxy luminosity for
the large relative errors for local galaxies. Fig. 2 is the
distribution of galaxy redshifts in our composite sample.
The first histogram with simple line is the redshift distribution
for the total galaxy sample. The second histogram is that for
galaxies with 20 colors. And the third histogram with filled area is
for galaxies with spectroscopic redshifts. Fig. 2 shows us directly the
contributions from these three sources.

%\clearpage
\begin{figure}[htbp]
\vspace{0.5cm}
\figurenum{1} \epsscale{1.1}
%\hspace{0.5cm}
%\rotatebox{-90}
{\plotone{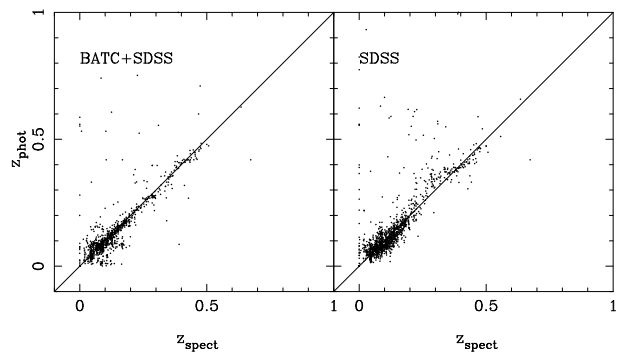}} \vspace{0.cm}
\caption{The comparison between photometric and spectroscopic redshifts estimated by 20 colors of BATC and SDSS
(left panel) and 5 colors of SDSS (right panel), respectively.}
\end{figure}

\begin{figure}[htbp]
\figurenum{2}
\vspace{0.5cm}
\epsscale{0.8}
%\hspace{2.0cm}
%\rotatebox{-90}
{\plotone{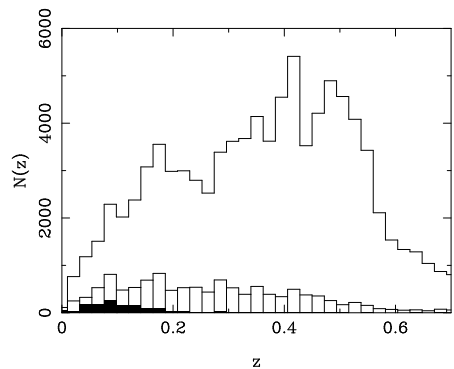}} \vspace{0.cm}
\caption{Redshift distributions for total galaxies (histogram with simple line), galaxies
with 20-color photometric redshifts and galaxies with spectroscopic redshifts (histogram with filled area).}
\end{figure}
%\clearpage

\subsection{$k$-correction}

By photometric redshifts we calculate absolute magnitudes in the restframe $r$-band of SDSS:
\begin{equation}
M_{r}=r-25-5\log d_{L}(z)-A_{r}-k(z).
\end{equation}
where $d_{L}(z)$ is the luminosity distance in Mpc, $A_{r}$ is the reddening extinction correction due to
intergalactic and interstellar dust scattering and absorption, and $k(z)$ is the $k$-correction due to
the shift of spectrum by redshift. $k$-correction is significant here since that the sampled redshifts
span to high redshifts. The common applied method is to estimate $k$-correction by galaxy color and
morphological type. In the advantage of photometric redshift fitting, we derive $k$-corrections
directly from the spectra templates. In photometric redshift fitting,
a best fit template can be achieved for every galaxy, which means the most semblable in the spectrum shape.
We modify slightly the code and
output the best fit template spectrum at $z$= 0. Assuming that the observed
spectrum has the same $k$-correction with the best fit template, the $k$-correction can be corrected
directly. The accuracy of this correction is investigated by simulation here.

By the procedure $make\_catalog$ \citep[see details in][]{xia02}, we build a catalog
of 1000 galaxies with random types of \citet{bruzual93} and redshifts in the range of $z<0.5$.
Totally 20 filters are used. The model
$k$-corrections are output directly from $make\_catalog$ and the estimated
corrections are output by $hyperz$. We, hence, can estimate the accuracy of the $k$-correction.
Fig. 3 shows us the distribution of the corrections with redshifts. The corrections $k(z)$
in the restframe $r$-band range from 0 to 2 mag. Fig. 4 shows the comparison
between model and estimated corrections, and the distribution of the deviations. The rms error of
$k$-correction is about 0.05 mag with a small offset of 0.01 mag.

Reddening extinction is obtained by step fitting in $hyperz$ \citep{xia02}. The reddening law
of \citet{allen76} for the Milky Way is adopted. The value of reddening correction is that best fitted
by the template spectrum. Here, we also assess the accuracy of the total correction of reddening
extinction $A_{r}$ and $k$-correction by simulation. It is found that the rms error of this total correction
is about 0.08 mag with a small offset of 0.01 mag.
In $\S$ 5 we will demonstrate by simulation that magnitude error plays
the most important role in the prediction of luminosity function and
this magnitude error will effect LF fit slightly.

%\clearpage
\begin{figure}[htbp]
\vspace{0.5cm}
\figurenum{3}
\epsscale{0.8}
%\hspace{2.0cm}
%\rotatebox{-90}
{\plotone{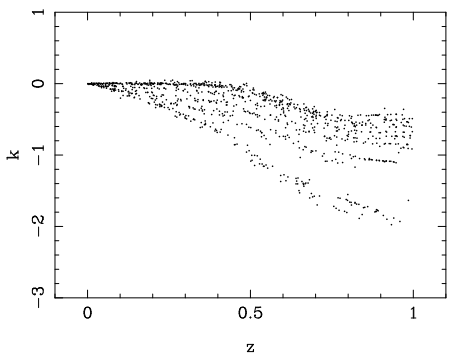}} \vspace{0.cm}
\caption{The distribution of simulated $k$-correction with redshifts.}
\end{figure}

\begin{figure}[htbp]
\vspace{0.5cm}
\figurenum{4}
\epsscale{1.2}
%\hspace{1.cm}
%\rotatebox{-90}
{\plotone{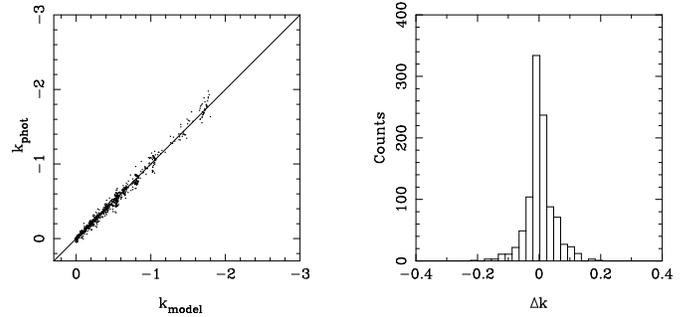}} \vspace{0.cm}
\caption{Assessment of the accuracy of $k$-correction and reddening extinction.
The left panel is the comparison between model correction and estimated correction.
The right panel is the distribution of the deviations.}
\end{figure}
%\clearpage

\section{EVOLUTION OF LUMINOSITY FUNCTION}

Luminosity function is the number density of galaxies as a function of luminosity.
To measure luminosity function, we adopt the approximation of Schechter function \citep{schechter76}:
\begin{equation}
\Phi(L)dL=\phi_{\ast}(\frac {L} {L_{\ast}})^{\alpha}\exp(-\frac {L} {L_{\ast}})d(\frac {L} {L_{\ast}}),
\end{equation}
expressed in the form of per unit absolut magnitude:
\begin{equation}
\Phi(M)dM=0.4\ln(10)\phi_{\ast}10^{-0.4(M-M_{\ast})(\alpha+1)}\exp[-10^{-0.4(M-M_{\ast})}]dM,
\end{equation}
where $M_{\ast}$ is the characteristic magnitude, i.e., the point at which the bright end cutoff sets in,
corresponding to $L_{\ast}$, $\alpha$ is the power-law slope of the faint end, and $\phi_{\ast}$
is the normalization constant. We search these parameters by $\chi^2$ minimization fitting. The fits
are performed over the magnitude range $-24<M_{i}<-16$ and the normalizations are done in
this magnitude range.

To correct the incompleteness arising from the selection effects of distance, the traditional
$1/V(M)$ method originally proposed by \citet{schmidt68} is implemented in this paper.
In the assumption that galaxies distribute homogeneously in comoving space, the correction function is $1/V(M)$,
where $V(M)$ is the maximum volume that determined by the maximum distance at which a galaxy with absolute
magnitude $M$ can be observed in the apparent magnitude limit. The comoving volume and luminosity distance
are calculated as that given by Hogg 1999:
\begin{equation}
dV_{C}=\frac{c}{H_{0}}\frac{d_{L}(z)^2}{(1+z)^2E(z)}d\Omega dz,
\end{equation}
\begin{equation}
d_{L}(z)=(1+z)\frac{c}{H_{0}}{\int_{0}^{z}\frac{dz'}{E(z')}},
E(z)=\sqrt{\Omega_{m}(1+z)^3+\Omega_{\Lambda}}
\end{equation}
The structure in small-scale and the possible evolution in number
density with redshfit can produce spurious estimate of local LF in this estimator. For the large size
of the data set it can partly make up the effect caused by inhomogeneity.
For the Malmquist Bias in distance estimates, it origins from that observation effect that
greater numbers of galaxies in the univers at greater distances and hence more will have been
scattered down from larger distances than up from smaller ones. We follow the method given by
\cite{lynden88} to correct the distance estimates. The correction formula is as below:
\begin{equation}
R=R_{e}\exp[(\alpha+{1\over2})\Delta^2],
\end{equation}
where $R_{e}$ is the estimated luminosity distance $D_{L}$, $\alpha=3$ for uniform distribution,
and $\Delta$ is the dispersion in $\ln{R_e}$.

To study the evolution of galaxy LF, we split our galaxy sample
into three redshift layers, $0.03<z<0.1$, $0.1<z<0.3$,
and $0.3<z<0.5$, with 5,289, 26,162, and 38,220 galaxies
respectively. Fig. 5 shows the LFs derived for
the three layers $0.03<z<0.1$ (filled triangles, solid line), $0.1<z<0.3$ (open
circles, dot-dashed line), and $0.3<z<0.5$ (filled circles, dotted line).
The error bars are the errors of poisson counts.
Measured parameters are given in Table 1 and the error contours are plotted
in the left panel of Fig. 6. As shown in Fig. 5 and
Table 1, the faint-end slope $\alpha$ steepens slightly from $-1.21\pm0.02$ to $-1.25\pm0.03$
and $-1.35\pm0.08$ with the increase of redshift.
The points appear some
discrepancy with Schechter function in bright end. The first
point exceeds than expectation. This may result from several
reasons. First, it may be due to the observation effects. For the
nearest galaxies, the uncertain proper motions can lead to large
errors in redshifts and hence overestimate the most luminous
galaxies. On the other hand, the amount of intermediate luminous
galaxies is much more than that of the most luminous galaxies in
the observation, the uncertainty of magnitude and photometric redshift,
therefore, can contaminate more intermediate redshift galaxies to
low redshifts and then bring to the excess of most luminous
galaxies.

%\clearpage
\begin{figure}[htbp]
\vspace{0.5cm}
\figurenum{5} \epsscale{1.}
%\hspace{4.0cm}
%\rotatebox{-90}
{\plotone{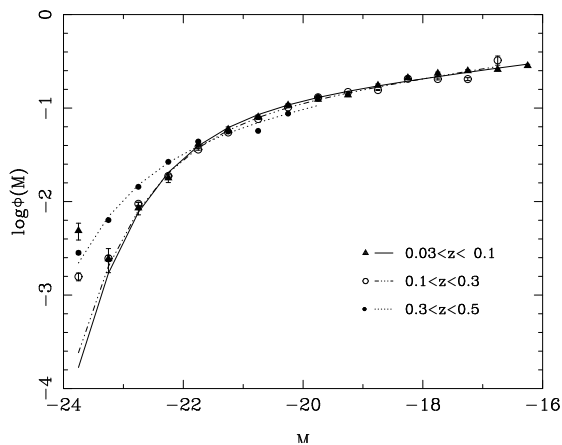}} \vspace{0.cm}
\caption{The luminosity functions fitted for the three redshift layers
$0.03<z<0.1$ (filled triangles, solid line), $0.1<z<0.3$
(open circles, dot-dashed line) and $0.3<z<0.5$ (filled circles, dotted line).}
\end{figure}
%\clearpage

\setcounter{table}{0}
\begin{table}[ht]
\caption[]{Shape parameters of LFs fitted to the three
redshift-binned subsamples and that fitted to high and low density
subsamples by density contrast $\delta\rho/\rho=30$.}
\vspace{0.3cm}
\begin{tabular}{ccccccccc}
\hline \hline
 & $\phi_{*}(\times10^{-2})$ & $M_{*}$ & $\alpha_{*}$ & $\chi^2$ \\
\hline
$0.03<z<0.1$/5289  & $10.16\pm0.08$ &  $-21.80\pm0.16$ & $-1.21\pm0.02$ &   0.43  \\
$0.1<z<0.3$/26162  & $8.56\pm0.13$ &  $-21.91\pm0.12$ & $-1.25\pm0.03$ &   2.59  \\
$0.3<z<0.5$/38220 & $4.42\pm0.17$ &  $-22.69\pm0.21$ & $-1.35\pm0.08$ &   29.77  \\
\hline
 high/54267 & $4.14\pm0.09$ &  $-22.64\pm0.18$ & $-1.37\pm0.04$ & 10.78  \\
  low/15404 & $8.83\pm0.05$ &  $-22.26\pm0.16$ & $-1.21\pm0.04$ &  2.92 \\
\hline
\hline
\end{tabular}
\end{table}

For local galaxies $0.03<z<0.1$, $\alpha=-1.21\pm0.02$ is
consistent with \citet{blanton01} $\alpha= -1.20\pm0.03$, which is
achieved by 11,275 galaxies complete to $r<17.6$ over 140 deg$^2$
in SDSS. \citet{blanton03} reevaluated galaxy luminosity function
at $z=0.1$ in the $^{0.1}r$ frame by a larger sample of
147,986 galaxies. A much flatter
faint-end slope $\alpha= -1.05\pm0.01$ was found. This large
difference is due to accounting for the evolution of luminosity
function in \citet{blanton03}.

From redshift layer $0.1<z<0.3$ to $0.3<z<0.5$, the estimated
faint-end slope is found to change from $\alpha= -1.25\pm0.03$
to $\alpha= -1.35\pm0.08$.
\citet{ellis96} constructed lminosity functions using a sample of
1,700 galaxies observed by Autofib Redshift Survey out to $z \sim
0.75$ and gave that $\alpha$ steepens from $-1.1$ to $-1.5$ with
redshift. Our results is in good agreement with this trend.
\citet{loveday04} studied the evolution of LF at $z<0.3$ by a
sample of 162,989 spectroscopic galaxies with magnitude limit of
$r<17.6$ in SDSS. The poorly constrained faint-end slope in redshift
slice $0.2<z<0.3$ is due to the incompleteness in high redshift
with a bright magnitude limit $r<17.6$. loveday04 investigated
the effect of different absolute magnitude ranges on the estimated faint-end
slope. It is found that the faint-end slope $\alpha$ changes from $-1.17$
to $-2.18$ if analysis of redshift slice $0.1<z<0.15$ is limited to
$M_{^{0.1}r}<-21.5$. This effect will propbably cause the overestimate
of faint-end slope for redshift layer $0.3<z<0.5$ in our sample:
basically there are no low-luminposity points to tie down the faint end.
Though not obvious of the incompleteness for the redshift slice
$0.1<z<0.3$ in Fig. 5, it could result from an incomplete sample, either.

The estimated characteristic magnitudes $M_{\ast}=-21.80\pm0.16$,
$M_{\ast}=-21.91\pm0.12$, and $M_{\ast}= -22.69\pm0.21$ for the
three redshift layers, are about 1 to 2 magnitude brighter
than previous results \citep{ellis96,blanton01,blanton03,loveday04}.
It is partly due to the different choice
of the hubble constant with $h= 0.75$ with others $h=1$.
Another reason comes from the effect of larger magnitude errors and photometric
redshift errors than spectroscopic samples. We will demonstrate
this effect in $\S$ 5.

%\clearpage
\begin{figure}[htbp]
\vspace{0.5cm}
\figurenum{6}
\epsscale{1.1}
%\hspace{3.0cm}
%\rotatebox{-90}
{\plotone{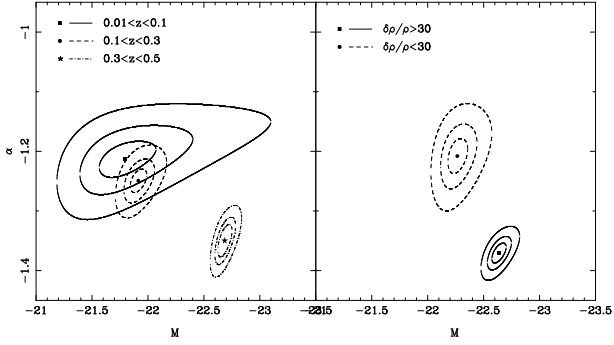}} \vspace{0.cm}
\caption{1-$\sigma$, 2-$\sigma$ and 3-$\sigma$ error contours for parameters
$M_{\ast}$ and $\alpha$. Left panel is that for the three redshift layers
$0.03<z<0.1$ (solid line), $0.1<z<0.3$ (dashed line) and $0.3<z<0.5$(dot-dashed line)
subsamples. Right panel is that for overdense (solid line) and underdense
(dot-dashed line) subsamples.}
\end{figure}
%\clearpage

\section{DEPENDENCE OF LUMINOSITY FUNCTION ON ENVIRONMENT}

To study the dependence of galaxy luminosity function on large-scale density environment, we subdivide
galaxies into high and low density two subsamples according to the density enhancements in three-dimensional
redshift space. There are many methods implemented in the literature, here we use the percolation
algorithm \citep{huchra82}. This algorithm can search isolate group and cluster.
The algorithm identifies every two galaxies by the projected separation $D_{12}$ and the line-of-sight
redshift separation $z_{12}$:
\begin{equation}
D_{12}=\sin(\theta /2)(z_{1}+z_{2})c/H_{0} < D_{L}(z_{1},z_{2},m_{1},m_{2}),
\end{equation}
\begin{equation}
z_{12}=|z_{1}-z_{2}| < z_{L}(z_{1},z_{2},m_{1},m_{2}),
\end{equation}
where $z_{1}$ and $z_{2}$ refer to the redshifts of the two galaxies in the pair, $\theta$ is their
angular separation, and $D_{L}$ and $z_{L}$
are scaled to account for the magnitude limit of the galaxy catalog. All pairs linked by a common
galaxy form a group. The limiting number density contrast is
\begin{equation}
\frac {\delta \rho} {\rho}=\frac {3} {4\pi D_{0}^{3}}[\int_{-\infty}^{M_{l}}\Phi(M)dM]^{-1}-1,
\end{equation}
where $\Phi(M)$ is luminosity function, $M_{l}$ is the faintest absolute magnitude for galaxies with magnitude
limit at fiducial distance. For a choosen density contrast and luminosity function (assumed without evolution
for the simplicity), the critical distance $D_{0}$ can be calculated. Taking into account the decrease of
galaxy numbers with increasing distance, link parameter $R$ can be calculated by:
\begin{equation}
R=[\frac {\int_{-\infty}^{M_{l}}\Phi(M)dM} {\int_{-\infty}^{M_{12}}\Phi(M)dM}]^{1/3},
\end{equation}
In $R$, $M_{12}$ is the faintest absolute magnitude for a galaxy with magnitude limit at the mean distance of
the two galaxies. Then $D_{L}=RD_{0}$, and $z_{L}=Rz_{0}$.
The magnitude limit here is the complete magnitude.
The fiducial redshift $z_{0}$ we choose is the minimum redshift
of the sample, $z_{f}=0.03$. $D_{0}$ is the fiducial distance corresponding to $z_{f}$.
$\Phi(M)$ is chosen as that measured by \citet{blanton01} for local galaxies by SDSS
commissioning data. The LF parameters are: $\phi_{\ast}=1.46\times10^{-2}h^3\rm{Mpc^{-3}}$, $M_{\ast}=-20.83$,
and $\alpha=-1.20$. By this criteria, we can obtained our subsamples by
$\delta\rho / \rho$. $\delta\rho / \rho = 30$ is chosen for our classification, with the
fiducial distance $D_{0}$ corresponding to about 0.35 $\rm{Mpc}$.
In this scale, the high density subsample includes group galaxies and cluster
galaxies (members larger than 5) and low density subsample contains field and void galaxies.
We obtain two subsamples with 54,267 and 15,404 galaxies, respectively.

Fig. 7 shows us the luminosity functions derived for subsamples classified by density
environment. The discrepancy in fits may be due to the same reason given above.
The best-fit Schechter parameters along with the number of galaxies considered in each density environment are
listed in Table 1. From Table 1 we can see that, with the enhancement of density, the faint-end slope
increases from $-1.21\pm0.04$ to $-1.37\pm0.04$ and the characteristic magnitude
brightens slightly from $-22.26\pm0.16$ to $-22.64\pm0.18$.
The right panel in Fig. 6 shows us the 1-$\sigma$ (68.3\% 2-parameter),
2-$\sigma$ (95\% 2-parameter) and 3-$\sigma$ $\chi^2$ contours in the $\alpha-M_{\ast}$ plane
for high and low density populations. The difference is significantly obvious at the confidence
level of 95\%. Since the curves in Fig. 7 are normalized in the magnitude range $-24<M_{i}<-16$,
the bright-end density of underdense populations
does not mean the numbers of luminous galaxies are larger than that in overdense regions.

%\clearpage
\begin{figure}[htbp]
\vspace{0.5cm}
\figurenum{7}
\epsscale{1.}
%\hspace{4.0cm}
%\rotatebox{-90}
{\plotone{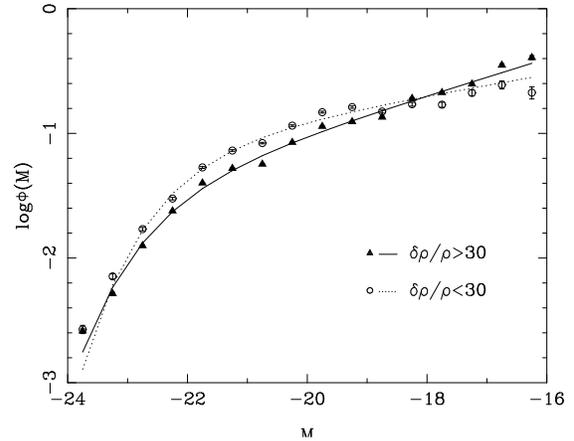}} \vspace{0.cm}
\caption{Luminosity functions fitted for overdense and underdense subsamples.}
\end{figure}
%\clearpage

\citet{croton04} indicates the luminosity function by density environment. They implement galaxies
observed by 2dFGRS with median survey depth is $z\approx0.11$. The local density contrast is determined
by $\delta_{8}$. The same tendency of $\alpha$ is given for regions of different density contrast.
In numerical simulations, Mo et al. 2004 studied the dependence of the galaxy luminosity function on large-scale
environment in hierarchical cosmology. The results predict that the characteristic luminosity,
$L^{\ast}$, increases moderately with density and the faint-end slope is quite independent of density in total.
$\alpha$ is virtually constant for late types and increases from $-1.3$ in underdense regions to $-1.8$
in overdense regions for early types. Although our samples are not split into subsamples by galaxy types,
the predictions are broadly in good agreement.
The steepen of the faint-end slope from underdense environment to overdense environment
is explained in \citet{tully02} by a process of photoionization of the IGM which suppressed dwarf galaxy formation.
Overdense regions typically collapse early and can form a dwarf galaxy before the epoch of reionization.
Underdense regions, however, collapse later and are thus subject to the photoionization suppression of cooling baryons.
The brighten of the characteristic magnitude in dense regions is consistent with the morphology-density relation
\citep{dressler80,binggeli88}, which is that the population in low density
subsamples is dominated by late types and cluster regions
have a relative excess of most luminous early-type galaxies. In structure evolution models, it is explained that,
the most dense regions of the universe will have collapsed earlier, have larger merger rate and will contain more massive
early type galaxies.

\section{DISCUSSION AND CONCLUSION}

The uncertainties of galaxy photometric redshifts are about from 0.01 to 0.04 in different photometric
magnitude uncertainties. This is the best accuracy achieved by multi-color photometric information.
This uncertainty, however, is much bigger than that of spectroscopic redshift, which can
be 0.0001. To calculate luminosity function by photometric redshifts, we need to know how the redshift
uncertainty and photometric magnitude uncertainty affect the shape of luminosity function.
To assess this effect, we perform a series of simulations to fit luminosity function for a galaxy
sample distributed as given luminosity function. 100 galaxy samples are created to evaluate the uncertainty of
fitted $M_{\ast}$ and $\alpha$. Each sample contains 5000 galaxies homogeneously distributed in the
1 deg$^2$ cone-shaped comoving space in the redshift range $z<0.6$. Gaussian distributed photometric
errors with $\Delta m$ = 0.05, 0.20, 0.50 are assumed. Gaussian distribution with a Gaussian kernel
of width ${\sigma}_z=\hat{\sigma}_z(1+z)$ for photometric redshift erros are adopted, where $\hat{\sigma}_z$ is the
redshift rms residual at zero redshift \citep{fernandez01,chen03}. $\hat{\sigma}_z$ = 0.05, 0.20, and 0.50
are assumed. Here we choose a given luminosity function with $M_{\ast}=-21.17$ and $\alpha=-1.26$.

Table 2 lists the estimated rms errors of $M_{\ast}$ and $\alpha$
for different photometric uncertainties and redshift uncertainties. We can find that,
the faint-end slope $\alpha$ become flatter and the characteristic magnitude $M_{\ast}$
become brighter for all uncertainties. It means that the shapes of luminosity function become flatter,
which is demonstrated intuitionisticly in Fig. 8. Fig. 8 shows us the luminosity function distribution
in Gaussian distributed photometric uncertainties of 0.05 and redshift uncertainties of 0.05 and 0.20.
Fig. 9 gives the distribution of $M_{\ast}$ and $\alpha$
in the 100 simulations with the same uncertainties as left panel in Fig. 9.
From Table 2, with the increase of redshift uncertainties, the characteristic magnitude $M_{\ast}$, is
determined with an offset around 0.2 mag brighter and an rms uncertainty about 0.1 mag, and $\alpha$ is determined
with an offset around 0.2 flatter and an rms uncertainty about 0.02.
The errors become larger with the increase of photometric and redshift uncertainties.
We can see by simulation that photometric errors and redshift errors are two major factors for
the measurement of luminosity function. This change trends of $M_{\ast}$ and $\alpha$ can explain
partly the difference between our estimates and previous results.

%\clearpage
\begin{figure}[htbp]
\vspace{0.5cm}
\figurenum{8}
\epsscale{1.2}
%\hspace{2.0cm}
%\rotatebox{-90}
{\plotone{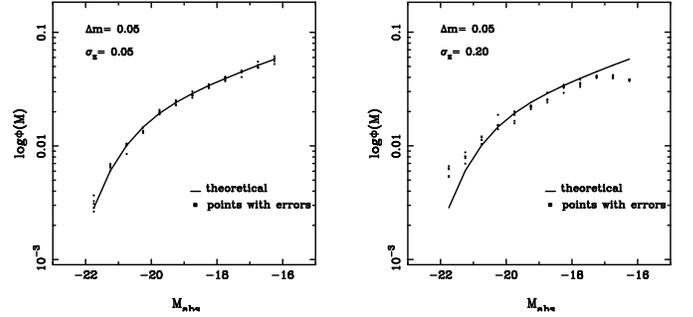}} \vspace{0.cm}
\caption{Luminosity function distribution of the 100 times simulations in the Gaussian distributed photometric
uncertainty of 0.05 and redshift uncertainty of 0.05 and 0.20.}
\end{figure}

\begin{figure}[htbp]
\vspace{0.5cm}
\figurenum{9}
\epsscale{1.2}
%\hspace{1.cm}
%\rotatebox{-90}
{\plotone{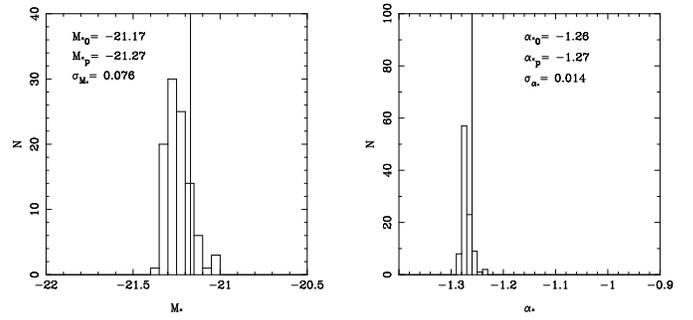}} \vspace{0.cm}
\caption{Distribution of fitted $M_{\ast}$ and $\alpha$ for the simulated
samples same as that of left panel of Fig. 8.}
\end{figure}
%\clearpage

\setcounter{table}{1}
\begin{table}[ht]
\caption[]{Simulated results of the accuracy of luminosity function parameters  
in different magnitude uncertaineties and redshift uncertainties. 
The given paramters of luminosity function are $M_{*}=-21.17$ and $\alpha_{*}=-1.26$.}
\vspace{0.3cm}
\begin{tabular}{cccccc}
\hline \hline 
$\sigma_{\rm {m}}$ & $\sigma_{\rm {z}}$ & $M_{\rm {*,fitted}}$ & $\sigma_{\rm {M_{*}}}$ 
& $\alpha_{\rm {*,fitted}}$ & $\sigma_{\rm {\alpha_{*}}}$ \\
\hline
0.05 & 0.05  &  -21.28 & 0.076 & -1.28 & 0.014 \\
0.05 & 0.20  &  -21.53 & 0.137 & -1.19 & 0.017 \\
0.05 & 0.50  &  -21.38 & 0.123 & -0.93 & 0.022 \\
0.20 & 0.05  &  -21.28 & 0.071 & -1.28 & 0.013 \\
0.20 & 0.20  &  -21.53 & 0.147 & -1.21 & 0.028 \\
0.20 & 0.50  &  -21.38 & 0.120 & -0.94 & 0.023 \\
0.50 & 0.05  &  -21.28 & 0.065 & -1.28 & 0.015 \\
0.50 & 0.20  &  -21.53 & 0.118 & -1.18 & 0.017 \\
0.50 & 0.50  &  -21.48 & 0.117 & -0.95 & 0.028 \\
\hline
\end{tabular}
\end{table}

We study the evolution of galaxy luminosity function in the
restfram $r$-band and the dependence of luminosity function on
density environment out to $z \sim 0.5$ by 69,671 galaxies
composed by BATC sky survey and SDSS sky survey.
We further the depth by photometric redshifts
and adopt the Schechter funtion as luminosity function model.
The evolution of galaxy luminosity
function is studied by three redshift layers $0.03<z<0.1$,
$0.1<z<0.3$ and $0.3<z<0.5$. We subdivide the density
environment by criteria of isodensity contour. The density
contrast is choosen to be $\delta\rho/\rho$= 30. The principle
conclusions are summarized as follows:

(i) By simulation we find that photometric uncertainty and redshift uncertainty are
two major factors that effect the measurement of luminosity function.
In photometric uncertainty of $\Delta m < 0.50$ and redshift uncertainty of
$\hat{\sigma_z}< 0.50$, the characteristic magnitude $M_{\ast}$, can be estimated 0.2 mag brighter
with a rms error about $\sigma_{M_{\ast}}$= 0.1; and the faint-end slope $\alpha$, can be recovered
with a factor 0.2 flatter and small uncertainty about $\sigma_{\alpha}$= 0.02.

(ii) There is slight evolution in the shape of galaxy luminosity function with observational depth.
We further the accurate measurement of galaxy luminosity function to $z<0.3$.
The faint-end slope steepens slightly from $-1.21$ to $-1.25$, $-1.35$ with the increase of redshift
from $0.03<z<0.1$ to $0.1<z<0.3$, $0.3<z<0.5$. The change of $\alpha$ is broadly consistent with
previous claims such as \citet{lilly95,ellis96}.

(iii) Luminosity function differs distinctly with density environment. The faint-end slope for high density
galaxies is steeper than that for low density galaxies. The value of $\alpha$ changes from $-1.21$ at
underdensity regions to $-1.37$ at high density regions. $M_{\ast}$ brightens from $-22.26$ to $-22.64$.

\acknowledgments
%{ACKNOWLEDGMENTS}

We would like to thank the referee Dr. Jon Loveday for his insightful comments and suggestions
that improved this paper significantly.
We would like to acknowledge very helpful discussions with Zhenyu Wu, Jiuli Li,
Zhonglue Wen, Tianmeng Zhang, Jianghua Wu, and Yuzhi Duan. And we would like to thank
Dr. David Burstin and Dr. Zuhui Fan for their helpful revision and suggestions over the paper.
This work has been supported by the Chinese National Nature Science Foundation, No. 10473012 and 10573020.

%\clearpage

%\input{table3.tex}

%\input{table4.tex}

\end{document}